# Anomalies of transport properties in antiferromagnetic YbMn$_2$Sb$_2$ compound

V.N. Nikiforov[a], V.V. Pryadun[a], A.V. Morozkin[b], V.Yu. Irkhin*[c]

[a]*Physical Department, Moscow State University, 119899, Russia*

[b]*Chemical Department, Moscow State University, 119899, Russia*

[c]*Institute of Metal Physics, 620990 Ekaterinburg, Russia*

*E-mail: valentin.irkhin@imp.uran.ru

The low-temperature transport properties (resistivity, thermal conductivity, thermoelectric power) are experimentally investigated for the compound YbMn$_2$Sb$_2$. Large Seebeck coefficient and nearly linear temperature dependence of resistivity are obtained. The thermal conductivity is large and has mainly phonon origin. Appreciable anomalies of transport characteristics near the antiferromagnetic transition point are found. Possible influence of pseudo-Kondo scattering is discussed.

## 1. Introduction

Ternary Yb-based intermetallic compounds YbT$_m$X$_n$ (T is transition metal, X = Si, Ge, Sb…) demonstrate a rich variety physical properties, including intermediate-valence and Kondo-latiice behavior, unusual magnetic ordering and anomalous transport properties [1]. In particular, magnetic properties of YbMn$_2$X$_2$ compounds have been investigated. The compound YbMn$_2$Si$_2$ exhibits a number of magnetic transitions: a collinear antiferromagnetic ordering of Mn moments at 526 K and an antiferromagnetic rearrangement of Mn sublattice below 30 K followed by the magnetic ordering of Yb$^{3+}$ moments at 1.5 K [2]. In the compound YbMn$_2$Ge$_2$ a planar antiferromagnetism changes to noncollinear-type ordering below 185 K, whereas no Yb sublattice magnetic ordering is observed down to 4.2 K [3]. XPS measurements show that itterbioum is in 3+ state in YbMn$_2$Si$_2$ and in 2+ state in YbMn$_2$Ge$_2$ [4].

Some of the Yb systems are promising thermoelectric materials due to high Seebeck coefficient and low thermal conductivity [5]. In particular, intermetallic compounds YbZn$_2$Sb$_2$ and YbZn$_{2-x}$Mn$_x$Sb$_2$ (YbZn$_2$Sb$_2$ and YbMn$_2$Sb$_2$ are isostructural, and the Mn substitution effectively lowers the thermal conductivity) have at high temperatures large values of figure of merit *ZT*, up to 0.61–0.65 for x=0.05–0.15 at 726 K [6].

At the same time, from the physical point of view, low-temperature transport properties are of interest to gain information about electronic structure. In the present work we study the transport properties of pure YbMn$_2$Sb$_2$. Although specific heat measurements performed down to 1.8 K in [7] did not yield a substantially enhanced value of the electronic specific heat, characteristic of mixed-valent of Kondo systems (the Sommerfeld coefficient γ = 29.7 mJ/mol K$^2$ is only moderately enhanced) we will find some non-trivial features. They are partly due to magnetism. The neutron diffraction investigation [8] showed that YbMn$_2$Sb$_2$ has antiferromagnetic ordering of the manganese atom moments below 120 K, no local moment being detected on the Yb sites (a situation of itinerant magnetism, similar to pure manganese).

In this present paper, we report the data on temperature dependences of resistivity, thermal conductivity and thermoelectric power, and discuss for the first time their anomalies at the Neel point.

## 2. Experimental

Ytterbium (pieces cut from ingot, with purity 99.9wt.%), manganese (small grains from a platelet previously surface-cleaned by $HNO_3$, with 99.99 wt.% purity) and antimony (grains, with 99.999 wt.% purity) were used as the starting components.

As a first synthesis step, the equiatomic binary alloy MnSb was prepared by induction melting of the elements in outgassed Ta crucibles, closed by welding under pure Ar, by heating up to 1250–1300 °C. Manganese antimonide was formed by nearly congruent melting, its formation temperature (840 °C) being relatively low (much lower than that the melting point of Mn metal, 1246 °C). At the second step, Yb and MnSb were mixed in the stoichiometric amounts, sealed again under Ar into an outgassed Ta crucible, and reacted by induction heating up to about 1300 °C. The crucibles were then sealed under vacuum in quartz tubes and annealed in a resistance furnace at 800 °C for 7 days; after annealing, they were air cooled. The quality of the polycrystalline alloys (MnSb and both arc and induction melted $YbMn_2Sb_2$ samples) was determined using X-ray powder diffraction and electron microscopy (SEM) equipped with EDX microprobe analysis.

Main phase (about 95% ) is trigonal $La_2O_2S$-type structure $YbMn_2Sb_2$, space group P3bm1 (N164), a = 4.5241(6) A, c = 7.4410(8) A, Yb (1a) (0, 0, 0), Mn (2e) (1/3, 2/3,0.6369(9)), Md (2e) (1/3 2/3, 0.2459(9)). The sample demonstrates weak ferromagnetism at room temperatures because of impurity of MnSb ($T_C$ =583 K) about 0.05 %.

The transport measurements were carried out in the temperature range of 2.0–380K by the Quantum Design Physical Property Measurement System. This enables simultaneous measurements of thermal conductivity, Seebeck coefficient and resistivity by monitoring both the temperature and voltage drop across a sample as a heat pulse is applied to one end. Measurements were performed under high vacuum (~$10^{-4}$ torr) using a four-probe lead configuration. The detailed information on the measuring technique is presented in Ref. [5].

## 3. Results

Electrical resistivity measured in the range of 2–320 K is metallic and varies linearly with temperature in a rather wide region (Fig.1).

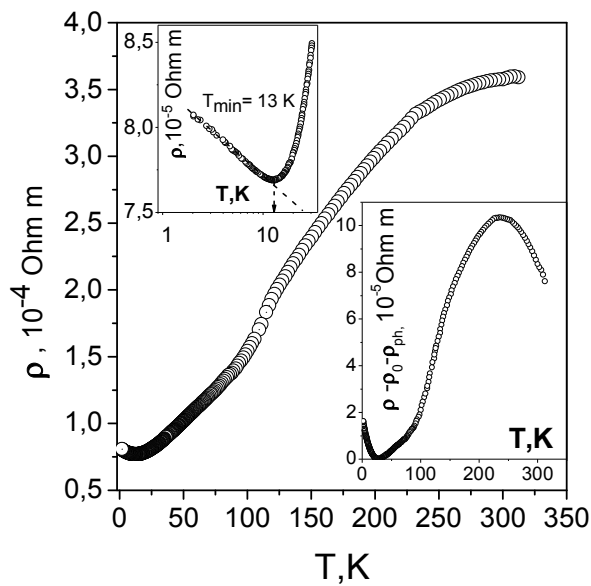

Fig.1 Temperature dependence of resistivity. The upper inset shows resistivity at low temperatures. The upper inset shows the dependence with residual and lattice resistivity, $\rho_0$ and $\rho_{ph}$, being subtracted

One can see the anomaly at the Neel point which is weaker than for compounds with "magnetic" Yb [1]. (Preliminary resistivity data [9] did not reveal the anomaly.) Such a behavior of resistivity near $T_N$ is observed in other antiferromagnetic metals (see Ref.[10,11]).

As the temperature is lowered, resistivity attains a minimum and further increases, the increase being described by a logarithmic law. The residual resistivity is rather large. Using the data of Ref.[7] on specific heat we can determine the coefficient at $T^3$-term as $\beta = 0.74$ mJ/mol K$^4$ and estimate the Debye temperature as $\theta_D = 140$ K. Then, by using the Bloch-Grüneisen formula, we can extract the lattice (phonon) contribution to resistivity (Fig.1). Thus magnetic resistivity demonstrates a high-temperature maximum. This may be attributed to anomalous Kondo-like scattering.

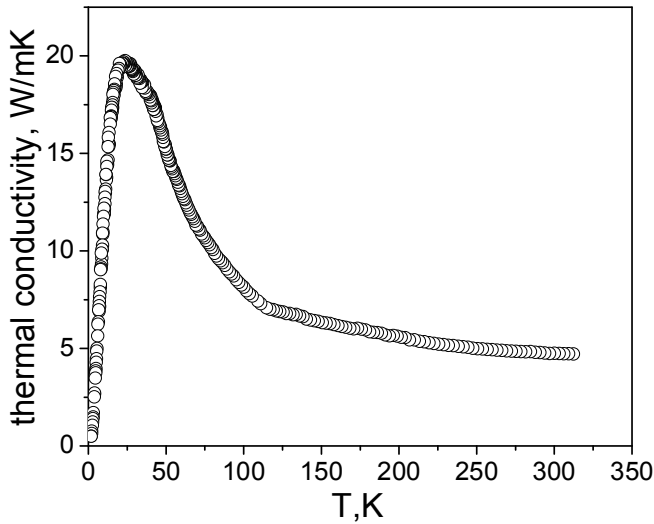

Fig. 2. Temperature dependence of thermal conductivity

Temperature dependence of the thermal conductivity $\kappa(T)$ is shown in Fig.2. Because of high resistivity of the sample, electron thermal conductivity is very small. Thus thermal conductivity has mainly phonon character and is considerably (by several times) larger in comparison with scuterrudite systems, where electron contribution dominates over lattice one or has the same order of magnitude (see, e.g., [3, 12]).

The dependence $\kappa(T)$ demonstrates a pronounced maximum at about 20K and has also a weak anomaly at the Neel temperature. It is difficult to explain this anomaly by pure electron contribution, so that a kind of spin-lattice interaction may be supposed.

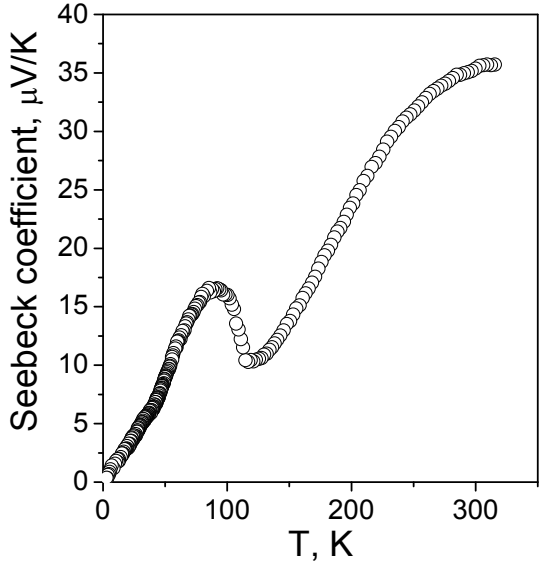

Fig.3. Temperature dependences of the Seebeck coefficient

Unusually strong anomaly in thermoelectric power (Fig.3) indicates considerable change in electronic structure near the Fermi level. Because of large lattice thermal conductivity, the Wiedemann-Franz law for the total κ does not hold and the figure of merit parameter *ZT* is small (of order of $10^{-4}$ at room temperatures). To improve thermoelectric properties, one needs also better conductivity.

### 4. Discussion and conclusions

A NFL-type (non-$T^2$) behavior of resistivity is also observed in a number of rare-earth and actinide systems [13]. Typically, a *T*-linear resistivity can be observed near the quantum critical point, as it takes place, e.g.., in YbRh$_2$Si$_2$ [14]. One of possible reasons of such a behavior in some region is scattering by antiferromagnetic fluctuations, especially in the case of nested Fermi surface [15]. Anomaly of electron spectrum at the Fermi surface also confirmed by anomalies of transport properties at $T_N$. However, a similar linear behavior was observed in YbPd$_2$Ge$_2$ (apparently non-magnetic) [1].

Upturn at very low *T* and the high-temperature maximum may be connected to a kind of pseudo-Kondo effect owing to two-level systems, e.g., with non-centrosymmetric scattering by Sb atoms, similar to the situation in IV-VI semiconductors [16] or Y$_3$Ir$_4$Ge$_{13}$ [17].

To conclude, we have investigated transport characteristics compound YbMn$_2$Sb$_2$ in a wide temperature region. We have found their appreciable anomalies (especially for thermoelectric power) near the antiferromagnetic transition point. Some features of resistivity are connected with possible pseudospin scattering.

The authors are grateful to Dr. R. Nirmala for discussion of the data on specific heat. This work is supported in part by the Programs of fundamental research of RAS Physical Division "Quantum macrophysics and nonlinear dynamics", project No. 12-T-2-1001 (Ural Branch) and of RAS Presidium "Quantum mesoscopic and disordered structures", project No. 12-P-2-1041.